\documentclass[12pt]{reportj}
\usepackage{deluxetablej}
\usepackage{hyperref} 
\makeatletter
\def\@to{to}
\makeatother
\usepackage{times}
\usepackage{graphicx}
\usepackage{tocloft}
\usepackage{fancyheadings}
\usepackage{hanging}
\usepackage{float}

\emergencystretch=\maxdimen
\usepackage{datetime}
\newdateformat{ddmonthyyyy}{\THEDAY\ \monthname[\THEMONTH]\ \THEYEAR}

\renewcommand\thesection{\arabic{section}}
\renewcommand\thesubsection{\thesection.\arabic{subsection}}

\def\ssection#1{\setcounter{subsection}{0} \refstepcounter{section} \section*{\hbox to \hsize{\large\bf \arabic{section}. #1\hfill }}\label{sec} \addcontentsline{toc}{section}{\arabic{section}. #1}}
\def\ssubsection#1{\setcounter{subsubsection}{0} \refstepcounter{subsection}\subsection*{\hbox to \hsize{\normalsize\bfseries\itshape \arabic{section}.\arabic{subsection} #1\hfill}}\label{subsec} \addcontentsline{toc}{subsection}{\arabic{section}.\arabic{subsection} #1}}
\def\ssubsubsection#1{\refstepcounter{subsubsection}\subsection*{\hbox to \hsize{\normalsize\it \arabic{section}.\arabic{subsection}.\arabic{subsubsection} #1\hfill}}\label{subsubsec} \addcontentsline{toc}{subsubsection}{\arabic{section}.\arabic{subsection}.\arabic{subsubsection} #1}}

\def\ssectionstar#1{\section*{\hbox to \hsize{\large\bf #1\hfill}} \addcontentsline{toc}{section}{#1}}
\def\ssubsectionstar#1{\subsection*{\hbox to \hsize{\normalsize\bfseries\itshape #1\hfill}} \addcontentsline{toc}{subsection}{#1}}
\def\ssubsubsectionstar#1{\subsection*{\hbox to \hsize{\normalsize\it  #1\hfill}} \addcontentsline{toc}{subsection}{#1}}

\renewcommand{\cftaftertoctitle}{%
\mbox{}\hfill{\normalfont Page}}
\begin{document}

~\\

\vspace{-2.4cm}
\noindent\includegraphics*[width=0.295\linewidth]{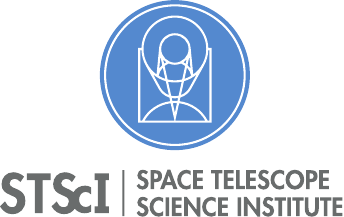}

\vspace{-0.4cm}

\begin{flushright}
    {\bf Instrument Science Report COS 2026-01(v1)}
    
    \vspace{1.1cm}
    
    {\bf\Huge Reference File Updates Following the Application of New Geometric Distortion and Walk Corrections I: Overview \par}

    \rule{0.25\linewidth}{0.5pt}
    
    \vspace{0.5cm}
    
    Nick Indriolo$^{1,2}$, W.~Fischer$^1$, E.~Frazer$^1$, D. French$^1$, S. Hasselquist$^1$, J.~Hernandez$^1$, C.~I.~Johnson$^1$, D. Kakkad$^{1,3}$, L.~P.~Miller$^1$, M.~Rafelski$^1$, R.~Sankrit$^1$
    \linebreak
    \newline
    \footnotesize{$^1$Space Telescope Science Institute, Baltimore, MD\\
    $^2$ AURA for European Space Agency, STScI, USA\\
    $^3$ Centre for Astrophysics Research, University of Hertfordshire, Hatfield AL10 9AB, UK\\}    
    \vspace{0.5cm}
    
     \ddmonthyyyy 2 February 2026
\end{flushright}

\vspace{0.1cm}

\noindent\rule{\linewidth}{1.0pt}
\noindent{\bf A{\footnotesize BSTRACT}}

{\it \noindent The calibration of Cosmic Origins Spectrograph (COS) data is performed by the data processing pipeline known as CalCOS. CalCOS utilizes several reference files that store information required for data processing (e.g., dispersion solutions), and these reference files can be updated for the purpose of improving calibrated data products. In 2024 a major effort to improve the far-ultraviolet (FUV) geometric distortion, delta-geometric, and walk corrections was completed. Those three corrections occur early in the CalCOS workflow and shift photon events into the corrected reference frame ((XCORR, YCORR) coordinate system) in which subsequent calibration steps---and their reference files---are defined. This change necessitated the re-derivation of all ``downstream'' reference files in use at the time. Here we summarize the effort to create new versions of 56 reference files applicable to lifetime positions 1 through 6 and intended for use with the new geometric distortion, delta geometric, and walk corrections. The primary outcome of this work is improved wavelength and flux calibrations for COS FUV spectra.
}

\vspace{-0.1cm}
\noindent\rule{\linewidth}{1.0pt}

\newpage
\renewcommand{\cftaftertoctitle}{\thispagestyle{fancy}}
\tableofcontents


\vspace{-0.3cm}
\ssection{Introduction}\label{sec:Introduction}

When processing Cosmic Origins Spectrograph (COS) data, the CalCOS pipeline begins by correcting data for instrumental effects (e.g., thermal distortion, geometric distortion, walk). Subsequent steps (e.g., wavelength calibration, spectral extraction, flux calibration) operate on data that have been corrected for those effects (Indriolo et al. 2025d), so their respective reference files must be derived from data that have been corrected in the same manner. The derivation of new geometric distortion and walk corrections for the FUV detector (summarized by Indriolo et al. 2025a) and application of their respective reference files results in photon events being moved to slightly different locations (i.e., being assigned to different pixels), effectively changing the coordinate system in which ``downstream'' reference files are defined. This has necessitated the effort to deliver new reference files for use with the updated geometric distortion and walk corrections, derived from data with those new corrections applied. In addition to the 5 reference files that define the new coordinate frame (BRFTAB, GEOFILE, DGEOFILE, XWLKFILE, and YWLKFILE), 56 downstream reference files required updates.

\lhead{}
\rhead{}
\cfoot{\rm {\hspace{-1.9cm} Instrument Science Report COS 2026-01(v1) Page \thepage}}

Reference files can be broken down into a few broad categories that perform different types of calibration actions. These include addressing data quality issues, extracting spectra, wavelength assignment, and flux assignment. Potential data quality issues are addressed by the BPIXTAB, GSAGTAB, and SPOTTAB files, which flag bad pixels, gain-sagged pixels, and pixels affected by hot spots, respectively.  Reference files associated with spectral extraction include XTRACTAB---used with the BOXCAR extraction method---and TWOZXTAB, PROFTAB, and TRACETAB---used with the TWOZONE extraction method. The LAMPTAB and DISPTAB reference files are used in the process of wavelength assignment, with the former aligning spectra to the frame where dispersion solutions are defined, and the latter applying those solutions to the spectra. Finally, flux calibration relies on the FLATFILE, FLUXTAB, and TDSTAB reference files. The FLATFILE removes large scale, low amplitude variations in the detector response, while the FLUXTAB converts photon count rates to fluxes (erg~s$^{-1}$~cm$^{-2}$~\AA$^{-1}$). The TDSTAB re-scales the count rate-to-flux conversion based on the measured decline in sensitivity as a function of time. More detailed descriptions of the reference files listed above can be found in the COS Data Handbook (Indriolo et al. 2025d). All of these files are defined either in the geometrically-corrected reference frame (XCORR, YCORR pixels) or in subsequent frames derived from that one (e.g., XFULL, YFULL, or wavelength), so they must also be updated. A full list of the new reference files is provided in the Appendix.

\ssection{Procedures}

The derivation and delivery of new reference files for COS is a routine procedure. Some files are updated regularly, while others are created/updated only when operations shift to a new lifetime position (LP) or new modes are commissioned. Files that apply to the entire detector (BRFTAB, BPIXTAB, GSAGTAB, and SPOTTAB) were re-derived first, followed by reference files applicable to individual LPs. We worked backwards in time (from LP6 to LP1) completing one LP before moving onto the next. Some reference files applied to LP1 and LP2 data are the same or use the same data in their derivation, so those two LPs were addressed in tandem. The last file to be re-derived was the TDSTAB, as it depends on flux-calibrated spectra at all LPs. 

\ssubsection{Files Applicable to the Whole Detector}

\ssubsubsection{BRFTAB}

The baseline reference frame table defines the locations of the stim pulses and the extent of the active area of the detector in the x and y directions.
While the detector edges would ideally be defined in a single coordinate system, the BRFTAB is currently used by CalCOS to check if photon events are within the active area via both (RAWX,RAWY) and (XCORR,YCORR) positions at different stages of pipeline processing to determine whether or not a given calibration step is applied. To ensure that all photon events in the active area are fully processed, the detector edges in the BRFTAB must be defined such that the full extent of the active areas in both (RAWX,RAWY) and (XCORR,YCORR) coordinate systems are encompassed. Changes made during the re-derivation of the geometric distortion and walk corrections moved the CORR positions of some photon events outside of the active area as defined in the previous BRFTAB, making it necessary to adjust the locations of the detector edges. A combination of nearly all FUV dark exposures taken from 2009 to 2023 was used to measure the left, right, top, and bottom edges of both detector segments in both the RAW and CORR coordinate frames. For a given edge, the less restrictive value between RAW and CORR coordinates was used to define that edge. Overall, the definitions of the detector edges in the BRFTAB were expanded slightly to ensure that calibration steps are applied to all photons recorded on the active area of detector. The updated baseline reference frame table is provided in the {\tt 97h1818gl\_brf.fits} reference file, and the HISTORY entry within that file provides more details about how the location of each detector edge was measured. This work was performed by S. Hasselquist.

\ssubsubsection{BPIXTAB}
Observations from programs 12676 (COS/FUV Characterization of Detector Effects), 15772 (Cycle 27 COS FUV Detector Gain Maps), 16472 (COS/FUV Gain Map and Aperture Placement at LP6), and 17248 (Cycle 30 COS FUV Detector Gain Maps) were used to provide semi-uniform illumination of the COS detector across the largest possible extent in both the dispersion and cross-dispersion directions. Data were processed using both old and new versions of the GEOFILE, DGEOFILE, XWLKFILE, and YWLKFILE reference files, and used to create old and new version 2D images of the detector that were compared side-by-side. Existing DQ regions were plotted on the old image to identify the detector features being flagged. Each feature was then located in the new image, and the DQ region responsible for flagging that feature was adjusted to account for any change in position caused by use of the new reference files. Four regions previously defined as low response regions (DQ=1024) were redefined to be very low response regions (DQ=16), and one very low response region was added to the FUVA segment. One pixel-out-of-bounds region (DQ=128) was added to the right edge of the FUVA segment to account for updates made to the BRFTAB reference file. The updated bad pixel table is provided in the {\tt 97h1816kl\_bpix.fits} reference file. This work was performed by D. Kakkad.

\ssubsubsection{GSAGTAB}
The GSAGTAB reference file contains 79 extensions, including the primary, with each data extension providing information about when and where a pixel sags (modal gain $<$ 3) as a function of detector segment high voltage (HV).  Specifically, each segment has an extension for HV = 0, 100, and then HV = 142 through 178.  Each extension contains a table with rows containing information about the MJD when a pixel first sagged, the lower left edge of the exclusion box in XCORR and YCORR coordinates, the exclusion box width and height, and the 8192 CalCOS flag. Since the geometric distortion and walk corrections affected both the coordinate system and the cumulative pulse height distributions, the GSAGTAB had to be regenerated from scratch.  After reprocessing all historical COS data with the updated geometric distortion and walk corrections, the COS team created weekly cumulative count images in which the pulse height distributions of every pixel per week were recorded.  A skewed Gaussian function was then fit to every pulse height distribution containing at least 25 counts at a given HV for a week, and the peak value (modal gain) was recorded.  The modal gain values were then fit with a linear function in 8 week increments, and on the date at which a pixel's fit equaled a value of 3 that pixel was permanently flagged as sagged for the measured HV and all lower HV values.  This information was then recorded in the final {\tt 97h18184l\_gsag.fits} reference file.  Note that the blue modes (G130M/1055 and G130M/1096) previously used a separate GSAGTAB; however, as part of the FUV recalibration effort two-zone extraction was added for the blue modes such that they no longer require a separate file. This work was performed by C.~I. Johnson.

\ssubsubsection{SPOTTAB}
The SPOTTAB reference file contains a single table with the FUV segment, start/end MJD of hot spot activity, the lower left XCORR/YCORR coordinates and width/height of the exclusion region, a DQ flag, and a comment labeling the hot spot of interest.  Unlike previous updates to the SPOTTAB, which relied on manual examinations of FUV exposures for discovery, the COS team enabled a new, automated method for detecting and tracking hot spots.  This method employs a ``Laplacian of Gaussian" method to automatically detect and characterize hot spots appearing in routine ($\sim$ biweekly) dark exposures.  When a spot is detected at a $>$ 5$\sigma$ level compared to the background, the region is flagged in the SPOTTAB for a time period covering the previous dark exposure and the subsequent dark exposure.  Testing showed that this method detected 100$\%$ of previously known hot spots, and also found several additional minor hot spots that were not previously flagged.  This information was recorded in the final {\tt 97h1816bl\_spot.fits} reference file. This work was performed by C.~I. Johnson.

\ssubsection{Files Specific to Lifetime Positions (LP1--LP6)}
The procedure for commissioning a new lifetime position involves the planning and execution of specific observations and the analysis of those observations to generate calibration reference files. Examples of this process are documented in Fischer et al. (2022) and Sankrit et al. (2023). In our case the requisite data already existed from the original commissioning activities, so our task was to identify the relevant exposures and utilize standard software routines to recreate the reference files, now in the updated coordinate frame. For a single LP, this includes creating new extraction files, wavelength calibration files, and flux calibration files. Three different data sets are used for this purpose. High signal-to-noise (S/N) observations of white dwarf (WD) calibrator stars are used to both define the extraction regions and derive the flux calibration. Observations of stars with many emission lines or interstellar absorption lines along the line of sight are used to define the dispersion solution. High S/N observations of the internal Pt-Ne wavelength calibration lamp tie the dispersion solution to a reference point.

\ssubsubsection{Workflow for LP6, LP5, LP4, and LP3}
At each LP we used the steps below to derive new reference files. All data were processed using the new GEOFILE, DGEOFILE, XWLKFILE, YWLKFILE, BRFTAB, BPIXTAB, GSAGTAB, and SPOTTAB. Because the procedure requires iteration, some reference files are derived multiple times. The creation of intermediate files are marked with a $^*$, while final versions are \underline{underlined}. When applicable, subscripts $b$ and $t$ denote files created using BOXCAR and TWOZONE extraction for PSA data, respectively.
\begin{enumerate}
\item{Use the calibrator WD observations to define the regions used in BOXCAR extraction (both for the PSA and WCA), and save these in a new \underline{XTRACTAB} reference file.}
\item{Use the calibrator WD observations and the new XTRACTAB to extract continuum spectra. Separate the dependencies on wavelength and detector pixels to create new FLATFILE$^*_b$ and FLUXTAB$^*_b$ reference files.}
\item{Use the Pt-Ne lamp observations and the new XTRACTAB and FLATFILE$_b$ to create new lamp template spectra, and save these in a new LAMPTAB$^*$ file.}
\item{Use the absorption/emission line observations and the new XTRACTAB, FLATFILE$_b$, FLUXTAB$_b$, and LAMPTAB to generate 1D spectra. Given the x-locations and known wavelengths of various lines, derive the dispersion solutions and save them to a new DISPTAB$^*_b$ reference file.}
\item{Process the calibrator WD data with the new FLATFILE$_b$, and use the resulting 2D images to create the profiles and traces used in TWOZONE extraction. Save these to new PROFTAB$^*$, TRACETAB$^*$, and TWOZXTAB$^*$ reference files.}
\item{Use the calibrator WD data and the new TWOZONE extraction files to repeat the analysis in step 2 and create new \underline{FLATFILE}$_t$ and \underline{FLUXTAB}$_t$ reference files.}
\item{Process the Pt-Ne lamp data with the new FLATFILE$_t$ and repeat the analysis in step 3 to create the final \underline{LAMPTAB} file.}
\item{Process the absorption/emission line observations using the new FLATFILE$_t$, FLUXTAB$_t$, LAMPTAB, and TWOZONE extraction files, and repeat the analysis in step 4 to derive dispersion solutions. These are saved in the final \underline{DISPTAB}$_t$.}
\item{Repeat step 5 using the newest FLATFILE$_t$ during processing, and create the final versions of \underline{PROFTAB}, \underline{TRACETAB}, and \underline{TWOZXTAB}. }
\item{Repeat step 6 using the ``final'' versions of all reference files to create a post-final FLATFILE$_t$. Compare the post-final FLATFILE$_t$ to the final FLATFILE$_t$ (created in step 6). If differences in the 1D profile are $\leq0.01$\%, then the iterative procedure is complete and the FLATFILE$_t$ from step 6 is considered the final version. If differences are $>0.01$\% then steps 9 and 10 are repeated until the FLATFILEs from successive iterations converge.}
\end{enumerate}

\ssubsubsection{Workflow for LP2 and LP1 Standard Modes} \label{sec_LP21_workflow}
The procedure for commissioning a new lifetime position was not yet standardized when operations shifted from LP1 to LP2 in 2012, and several file types that were originally thought to be applicable to both LPs are now recreated specifically for each new LP. Further improvements to LP1 and LP2 calibration over time resulted in a situation in which some files apply only to LP1, some only to LP2, and some to both LP1 and LP2 (denoted with subscripts ``1'', ``2'', and ``1,2'', respectively, below). Because the reference files for LP1 and LP2 are intertwined, and because standard mode observations at LP1 and LP2 only use BOXCAR extraction, these LPs required a different workflow. The decision to use only LP1 or only LP2 data to create a reference file that applies to both LPs was based entirely on how the existing versions of the reference files were created.
\begin{enumerate}
\item{Use calibrator WD observations at LP1 to define the PSA and WCA BOXCAR extraction regions at LP1, and save these in a new \underline{XTRACTAB}$_1$ file.} 
\item{Use calibrator WD observations at LP2 to define the PSA and WCA BOXCAR extraction regions at LP2, and save these in a new \underline{XTRACTAB}$_2$ file.}
\item{Use calibrator WD observations at LP2 and the new XTRACTAB$_2$ to extract continuum spectra. Separate the dependencies on wavelength and detector pixels to create new \underline{FLATFILE}$_{1,2}$ and \underline{FLUXTAB}$_2$ reference files.}
\item{Use calibrator WD observations at LP1 and the new XTRACTAB$_1$ to extract continuum spectra. Taking the FLATFILE$_{1,2}$ made in step 3 to be fixed, derive the sensitivity dependence on wavelength and save the result in a new \underline{FLUXTAB}$_1$ reference file.}
\item{Use the Pt-Ne lamp observations at LP1 and the new XTRACTAB$_1$ and FLATFILE$_{1,2}$ to create new lamp template spectra, and save these in a new \underline{LAMPTAB}$_{1,2}$ reference file.}
\item{Process emission/absorption line data at LP1 from several Calibration (Cal) and General Observer (GO) programs using the new XTRACTAB$_1$, FLATFILE$_{1,2}$, FLUXTAB$_1$, and LAMPTAB$_{1,2}$ to generate 1D spectra. Derive the dispersion solutions and save them to a new \underline{DISPTAB}$_1$ reference file.}
\item{Process emission/absorption line data at LP2 from several Cal and GO programs using the new XTRACTAB$_2$, FLATFILE$_{1,2}$, FLUXTAB$_2$, and LAMPTAB$_{1,2}$ to generate 1D spectra. Derive the dispersion solutions and save them to a new \underline{DISPTAB}$_2$ reference file.}
\end{enumerate}

\ssubsubsection{Workflow for G130M/1055 and G130M/1096 Modes at LP2}
Often referred to as the ``blue modes'', G130M/1055 and G130M/1096 were originally commissioned at LP2, and remained there until moving to LP7 at the start of HST Cycle 33. Because these modes have wide cross-dispersion profiles that overlap with heavily gain-sagged regions at LP1, extraction using the BOXCAR algorithm relied on a special GSAGTAB that only applied to these modes. We decided to retire this special GSAGTAB and instead implement the TWOZONE extraction method for just the blue modes at LP2. In some cases reference files apply only to the blue modes, while in other cases the blue mode entries appear within standard reference files. We specify which case applies to each file only when the final version of a reference file is generated in the workflow below. Pre-existing ``final'' reference files (i.e., files created during the workflow described in Section \ref{sec_LP21_workflow}) that are updated with new G130M/1055 and G130M/1096 entries are denote by \textit{italics}, while newly created final files that only apply to the blue modes are \underline{underlined}.
\begin{enumerate}
\item{Use calibrator WD observations to define the PSA and WCA BOXCAR extraction regions, and save these into the existing \textit{XTRACTAB}$_2$ file.} 
\item{Use calibrator WD observations and the updated XTRACTAB$_2$, relaxing constraints for gain sagged pixels, to create continuum spectra. Separate the dependencies on wavelength and detector pixels to create blue mode specific FLATFILE$^*_b$ and FLUXTAB$^*_b$ reference files.}
\item{Using a large number of Cal and GO observations at 1055 and 1096, extract, align, and combine WCA spectra to produce high S/N lamp template spectra. Save the results within a new LAMPTAB$^*$ file.}
\item{Use absorption line observations and the new XTRACTAB$_2$, LAMPTAB, and blue mode specific FLATFILE$_b$ and FLUXTAB$_b$ to generate 1D spectra. Derive the dispersion solutions and save them within a new DISPTAB$^*$ file.}
\item{Process the calibrator WD data with the new blue mode specific FLATFILE$_b$, and use the resulting 2D images to create the profiles and traces used in TWOZONE extraction. Save these to new PROFTAB$^*$, TRACETAB$^*$, and TWOZXTAB$^*$ reference files.}
\item{Use the calibrator WD data and the new TWOZONE extraction files to repeat the analysis in step 2. The detector response is saved into a new blue mode specific \underline{FLATFILE}$_t$, while the sensitivity curves are saved into the existing \textit{FLUXTAB}$_2$ file.}
\item{Process the Cal and GO data with the new FLATFILE$_t$ and repeat the analysis in step 3 to create new 1055 and 1096 lamp template spectra. Save the results into the existing \textit{LAMPTAB}$_{1,2}$ file.}
\item{Process the absorption line observations using the new FLATFILE$_t$, updated FLUXTAB$_2$ and LAMPTAB$_{1,2}$, and new TWOZONE extraction files, and repeat the analysis in step 4 to derive dispersion solutions. Save these into the existing \textit{DISPTAB}$_2$.}
\item{Repeat step 5 using the newest FLATFILE$_t$ during processing, and create the final versions of \underline{PROFTAB}, \underline{TRACETAB}, and \underline{TWOZXTAB}.}
\item{Repeat step 6 using the ``final'' versions of all reference files to create a post-final blue mode specific FLATFILE$_t$. Compare the post-final FLATFILE$_t$ to the final FLATFILE$_t$ (created in step 6). If differences in the 1D profile are $\leq0.01$\%, then the iterative procedure is complete and the FLATFILE$_t$ from step 6 is considered the final version. If differences are $>0.01$\% then steps 9 and 10 are repeated until the FLATFILEs from successive iterations converge.}
\end{enumerate}

\ssubsection{Files with Time Dependence}
The only reference file with time dependence is the TDSTAB, which contains the time-dependent sensitivity correction. Previously, there were two separate FUV TDSTAB files with different use-after dates. The file applied to data before 2017-10-02 had dependencies on segment and grating, while the file applied to data on or after 2017-10-02 included an additional dependence on cenwave. To better characterize the time dependent sensitivity for early COS observations, we decided to add cenwave dependence to the TDSTAB for all observations, consolidating the two reference files into one.

Re-derivation of the TDSTAB file required the analysis of calibrator WD observations over the entire history of COS on-orbit operations. Using all of the new reference files created in the workflows above, the relevant data were re-processed, and spectra were analyzed to track the changing sensitivity of the COS FUV modes as a function of time. A detailed description of this effort is provided in Hernandez et al. (2025).

Upon creation of the new TDSTAB file, one final step is required to complete the re-derivation of reference files; all FLUXTAB reference files must be updated to account for the new TDSTAB. This is the standard procedure whenever a new TDSTAB is delivered, and so all of the aforementioned ``final'' FLUXTAB reference files are replaced in this step. This marks the end of the effort required to create new reference files.

\ssubsection{Summary of Documentation}
The previously described workflows only give a broad sense of the procedures used in re-deriving reference files at each LP. In practice, the creation of new reference files involves many nuances and decisions that may change from one LP to the next for a variety of reasons. Detailed descriptions of the methods used to create the new reference files are provided in the series of Instrument Science Reports (ISRs) listed in Table \ref{tbl_isrs}.

\begin{deluxetable}{lll}
\tablewidth{0pt}
\tablecaption{Documentation of the FUV Recalibration Effort\label{tbl_isrs}}
\tabletypesize{\footnotesize}
\tablehead{\colhead{Document} &  \colhead{Citation} & \colhead{File Type(s) / Description}
}
\startdata
COS ISR 2025-07 & Indriolo et al. (2025a) & Geo/Walk Project Summary \\
COS ISR 2025-08 & Indriolo et al. (2025b) & DGEOFILE \\
COS ISR 2025-09 & Kakkad et al. (2025) & GEOFILE \\
COS ISR 2025-10 & Hasselquist et al. (2025a) & XWLKFILE \\
COS ISR 2025-11 & Hasselquist et al. (2025b) & YWLKFILE \\
COS ISR 2026-NN & French et al. (2026a) & Validation Testing  \\
\hline
COS ISR 2026-01 & Indriolo et al. (2026) & FUV Recalibration Summary (this document) \\
COS ISR 2025-19 & Hernandez et al. (2025) & TDSTAB, FLUXTAB\tablenotemark{a} \\
COS ISR 2025-15 & Indriolo et al. (2025c) & LAMPTAB \\
COS ISR 2026-NN & Hasselquist et al. (2026) & XTRACTAB, PROFTAB, TRACETAB, TWOZXTAB \\
COS ISR 2026-NN & Miller et al. (2026) & FLATFILE, FLUXTAB\tablenotemark{b} \\
COS ISR 2026-NN & French et al. (2026b) & DISPTAB 
\enddata
\tablecomments{Documents in the upper portion of the table describe the effort to derive the new geometric distortion, delta-geometric, and walk corrections. Documents in the lower portion of the table describe the effort to re-derive downstream reference files following completion of the work described in the upper portion of the table. An entry of ``NN'' within the document name denotes a work in preparation.}
\tablenotetext{a}{These are the final versions of the FLUXTAB reference files that are created after deriving the time-dependent sensitivity correction.}
\tablenotetext{b}{These are the near-final versions of the FLUXTAB reference files that are used when deriving the time-dependent sensitivity correction.}
\end{deluxetable}

\ssection{Validation and Outcome}

Before delivery to operations, all newly created reference files underwent a series of tests to confirm that (1) they work with CalCOS, and (2) they provide data products that are equivalent to or better than those produced by the previous set of reference files. The first test is simple, and was completed as part of the general workflow any time the new reference files were used during the calibration process. The second test is more complicated, and involved a comparison of newly created data products to fully calibrated data products that were created using the old set of reference files. 
The specific tests performed were different depending on the reference file(s) in question and are beyond the scope of this document. Detailed descriptions of the validation tests can be found in the ISRs listed in Table \ref{tbl_isrs}.

Application of the new set of reference files produces data products with improved wavelength and flux calibrations compared to data products that used the old reference files. The exact degree of improvement in any given spectrum varies with observing mode and location on the detector, but some general statements about the improvements are as follows:
\begin{enumerate}
\item{Accuracy of the overall wavelength assignment has improved from about 1/2 of a resolution element (3 pixels) to about 1/4 of a resolution element (1.5 pixels).}
\item{The maximum systematic offset between wavelengths assigned to features at different locations on the detector has been reduced from about 20 km~s$^{-1}$ to about 10 km~s$^{-1}$ for medium resolution modes.}
\item{Artificial flux spikes and deficits that had been caused by the old geometric distortion correction near the edges of the detector have been removed.}
\item{Errors in the absolute flux calibration for wavelengths recorded at the edges of the detector have been reduced to well within the 5\% requirement for the entire history of COS observations.}
\end{enumerate}
Various figures that demonstrate these improvements can be found within the documents listed in  Table \ref{tbl_isrs}.

\ssection{Summary}

Updates to the geometric distortion, delta-geometric, X-walk, and Y-walk corrections were made to improve COS FUV data products. Because these updates changed the coordinate system in which many downstream reference files are defined, the re-derivation of 56 reference files used for calibration at LP1 through LP6 was required before the new corrections could be implemented for general use. We have described the overall workflow that was used to create these reference files. All of the files were delivered to the HST Calibration Reference Data System on 2025 July 18 and are listed in the Appendix.



\vspace{-0.3cm}
\ssectionstar{Change History for COS ISR 2026-01}\label{sec:History}
\vspace{-0.3cm}
Version 1: \ddmonthyyyy 2 February 2026 

\vspace{-0.3cm}
\ssectionstar{References}\label{sec:References}
\vspace{-0.3cm}

\begin{hangparas}{.25in}{1}
Fischer, W. J., et al. 2022, COS ISR 2022-03, {\it Summary of COS Calibration for the Lifetime Position 5 Era}

French, D., et al. 2026a, COS ISR 2026-NN {\it Testing Metrics for Improvements to the COS FUV Geometric Distortion and Walk Corrections} (In Preparation)

French, D., et al. 2026b, COS ISR 2026-NN, {\it Reference File Updates Following the Application of New Geometric Distortion and Walk Corrections VI: DISPTAB} (In Preparation)

Hasselquist, S., et al. 2025a, COS ISR 2025-10, {\it Determining X-Walk Corrections for the COS FUV Detector}

Hasselquist, S., et al. 2025b, COS ISR 2025-11, {\it Determining Y-Walk Corrections for the COS FUV Detector}

Hasselquist, S., et al. 2026, COS ISR 2026-NN, {\it Reference File Updates Following the Application of New Geometric Distortion and Walk Corrections IV: extraction} (In Preparation)

Hernandez, J., et al. 2025, COS ISR 2025-19, {Reference File Updates Following the Application of New Geometric Distortion and Walk Corrections II: TDSTAB and HVDSCORR}

Indriolo, N., et al. 2025a, COS ISR 2025-07, {\it An Overview of Improvements to the COS FUV Geometric Distortion and Walk Corrections}

Indriolo, N., et al. 2025b, COS ISR 2025-08, {\it Measurement and Implementation of a Delta-Geometric Correction for the COS FUV Detector}

Indriolo, N., et al. 2025c, COS ISR 2025-15, {\it Reference File Updates Following the Application of New Geometric Distortion and Walk Corrections III: LAMPTAB}

Indriolo, N., et al. 2025d, {\it COS Data Handbook}, Version 6.0, (Baltimore: STScI).

Kakkad, D., et al. 2025, COS ISR 2025-09, {\it A Revised Geometric Distortion Correction for the Far-Ultraviolet Detector of the Cosmic Origins Spectrograph}

Miller, L., et al. 2026, COS ISR 2026-NN, {\it Reference File Updates Following the Application of New Geometric Distortion and Walk Corrections V: Flats and Fluxes} (In Preparation)



Sankrit, R., et al. 2023, COS ISR 2023-25, {\it Overview of COS Lifetime Position 6 Calibration}


\end{hangparas}

\ssectionstar{Appendix A}\label{sec:Appendix}

\begin{deluxetable}{cccccc}
\tablewidth{0pt}
\tablecaption{Reference Files Delivered for FUV Recalibration Effort\label{tbl_reffiles}}
\tabletypesize{\footnotesize}
\tablehead{\colhead{File Type} &  \colhead{File Name} & \colhead{LPs} & \colhead{Use After} & \colhead{Grating} & \colhead{Cenwave}
}
\startdata
BRFTAB & {\tt 97h1818gl\_brf.fits} & ... & 1996-10-01 & ... & ... \\
GEOFILE & {\tt 97h1817fl\_geo.fits} & ... & 2003-10-10 & ... & ... \\
DGEOFILE & {\tt 97h1817bl\_dgeo.fits} & ... & 2003-10-10 & ... & ... \\
XWLKFILE & {\tt 97h1816dl\_xwalk.fits} & ... & 2009-05-11 & ... & ... \\
YWLKFILE & {\tt 97h1817el\_ywalk.fits} & ... & 2009-05-11 & ... & ... \\
\hline
BPIXTAB & {\tt 97h1816kl\_bpix.fits} & ... & 2009-05-11 & ... & ... \\
DISPTAB & {\tt 97h1818ll\_disp.fits} & $-1$,1 & 2009-06-08 & ... & ... \\
DISPTAB & {\tt 97h1816gl\_disp.fits} & 2 & 2009-08-17 & ... & ... \\
DISPTAB & {\tt 97h1818ol\_disp.fits} & 3 & 2009-08-17 & ... & ... \\
DISPTAB & {\tt 97h1820ql\_disp.fits} & 4 & 2009-08-17 & ... & ... \\
DISPTAB & {\tt 97h1820cl\_disp.fits} & 5 & 2009-08-17 & ... & ... \\
DISPTAB & {\tt 97h1818bl\_disp.fits} & 6 & 2009-08-17 & ... & ... \\
FLATFILE & {\tt 97h18213l\_flat.fits} & $-1$,1,2 & 1996-10-01 & G130M & ... \\
FLATFILE & {\tt 97h18199l\_flat.fits} & $-1$,1,2 & 1996-10-01 & G140L & ... \\
FLATFILE & {\tt 97h18173l\_flat.fits} & $-1$,1,2 & 1996-10-01 & G160M & ... \\
FLATFILE & {\tt 97h1816ll\_flat.fits} & 2 & 1996-10-01 & G130M & 1055, 1096 \\
FLATFILE & {\tt 97h1818il\_flat.fits} & 3 & 1996-10-01 & G130M & ... \\
FLATFILE & {\tt 97h1818tl\_flat.fits} & 3 & 1996-10-01 & G140L & ... \\
FLATFILE & {\tt 97h1819hl\_flat.fits} & 3 & 1996-10-01 & G160M & ... \\
FLATFILE & {\tt 97h18178l\_flat.fits} & 4 & 1996-10-01 & G130M & ... \\
FLATFILE & {\tt 97h1817jl\_flat.fits} & 4 & 1996-10-01 & G140L & ... \\
FLATFILE & {\tt 97h1816ql\_flat.fits} & 4 & 1996-10-01 & G160M & ... \\
FLATFILE & {\tt 97h18193l\_flat.fits} & 5 & 2009-05-11 & G130M & ... \\
FLATFILE & {\tt 97h1818ql\_flat.fits} & 6 & 1996-10-01 & G160M & ... \\
FLUXTAB & {\tt 97h1819fl\_phot.fits} & $-1$,1 & 2009-05-11 & ... & ... \\
FLUXTAB & {\tt 97h18182l\_phot.fits} & 2 & 2009-05-11 & ... & ... \\
FLUXTAB & {\tt 97h1818ml\_phot.fits} & 3 & 2009-05-11 & ... & ... \\
FLUXTAB & {\tt 97h18177l\_phot.fits} & 4 & 2009-05-11 & ... & ... \\
FLUXTAB & {\tt 97h18172l\_phot.fits} & 5 & 2009-05-11 & ... & ... \\
FLUXTAB & {\tt 97h1817il\_phot.fits} & 6 & 2009-05-11 & ... & ... \\
GSAGTAB & {\tt 97h18184l\_gsag.fits}  & ... & 2009-05-11 & ... & ... \\
HVDSTAB & {\tt 97h1819cl\_hvds.fits}  & ... & 2009-05-11 & ... & ... \\
LAMPTAB & {\tt 97h18198l\_lamp.fits}  & $-1$,1,2 & 2009-06-08 & ... & ... \\
LAMPTAB & {\tt 97h1816il\_lamp.fits}  & $-1$,1,2 & 2009-08-17 & ... & ... \\
LAMPTAB & {\tt 97h18196l\_lamp.fits}  & 3 & 2009-08-17 & ... & ... \\
LAMPTAB & {\tt 97h1819el\_lamp.fits}  & 4 & 2009-08-17 & ... & ... \\
LAMPTAB & {\tt 97h1818cl\_lamp.fits}  & 5 & 2009-08-17 & ... & ... \\
LAMPTAB & {\tt 97h18210l\_lamp.fits}  & 6 & 2009-08-17 & ... & ... \\
PROFTAB &  {\tt 97h1817ml\_profile.fits}\tablenotemark{a} & 2 & 2009-05-11 & G130M & 1055, 1096 \\
PROFTAB &  {\tt 97h1819ll\_profile.fits} & 3 & 2009-05-11 & ... & ... \\
PROFTAB &  {\tt 97h1820el\_profile.fits} & 4 & 2009-05-11 & ... & ... \\
PROFTAB & {\tt 97h18201l\_profile.fits}  & 5 & 2009-05-11 & ... & ... \\
PROFTAB & {\tt 97h18162l\_profile.fits}  & 6 & 2009-05-11 & ... & ... \\
SPOTTAB & {\tt 97h1816bl\_spot.fits}  & ... & 2009-05-11 & ... & ... \\
TDSTAB & {\tt 97h18216l\_tds.fits}  & ... & 2009-05-11 & ... & ... \\
TRACETAB &  {\tt 97h1816ol\_trace.fits}\tablenotemark{a} & 2 & 2009-05-11 & G130M & 1055, 1096 \\
TRACETAB & {\tt 97h18209l\_trace.fits}  & 3 & 2009-05-11 & ... & ... \\
TRACETAB & {\tt 97h1816el\_trace.fits}  & 4 & 2009-05-11 & ... & ... \\
TRACETAB & {\tt 97h18170l\_trace.fits}  & 5 & 2009-05-11 & ... & ... \\
TRACETAB & {\tt 97h18189l\_trace.fits}  & 6 & 2009-05-11 & ... & ... \\
TWOZXTAB &  {\tt 97h1818fl\_2zx.fits}\tablenotemark{a} & 2 & 2009-05-11 & G130M & 1055, 1096 \\
TWOZXTAB & {\tt 97h18188l\_2zx.fits}  & 3 & 2009-05-11 & ... & ... \\
TWOZXTAB & {\tt 97h18212l\_2zx.fits}  & 4 & 2009-05-11 & ... & ... \\
TWOZXTAB & {\tt 97h1820tl\_2zx.fits}  & 5 & 2009-05-11 & ... & ... \\
TWOZXTAB & {\tt 97h1820bl\_2zx.fits}  & 6 & 2009-05-11 & ... & ... \\
XTRACTAB & {\tt 97h1816tl\_1dx.fits}  & $-1$,1 & 2009-05-11 & ... & ... \\
XTRACTAB & {\tt 97h1818el\_1dx.fits}  & 2 & 2009-05-11 & ... & ... \\
XTRACTAB & {\tt 97h1818pl\_1dx.fits}  & 3 & 2009-05-11 & ... & ... \\
XTRACTAB & {\tt 97h1816hl\_1dx.fits}  & 4 & 2009-05-11 & ... & ... \\
XTRACTAB & {\tt 97h1820rl\_1dx.fits}  & 5 & 2009-05-11 & ... & ... \\
XTRACTAB & {\tt 97h1819kl\_1dx.fits}  & 6 & 2009-05-11 & ... & ... \\
\enddata
\tablecomments{An entry of ... indicates that the reference file has no dependence on this parameter. Files in the top section of the table apply the new geometric distortion and walk corrections. Files in the bottom section of the table extract and calibrate spectra from observations. An LP entry of $-1$ indicates that this file is applied to observations taken at any cross-dispersion location on the detector that does not correspond to a defined lifetime position.}
\tablenotetext{a}{This file is used only for TWOZONE extraction of G130M/1055 and G130M/1096 data at LP2.}
\end{deluxetable}

\end{document}